\documentclass[twoside,twocolumn]{article}

\usepackage{blindtext} 

\usepackage[sc]{mathpazo} 
\usepackage[scale=0.93]{tgpagella} 
\usepackage[T1]{fontenc} 
\usepackage{microtype} 

\usepackage[english]{babel} 

\usepackage[hmarginratio=1:1,top=32mm,columnsep=20pt]{geometry} 
\usepackage[hang, small,labelfont=bf,up,textfont=it,up]{caption} 
\usepackage{float}
\usepackage{booktabs} 

\usepackage{lettrine} 

\usepackage{enumitem} 
\setlist[itemize]{noitemsep} 

\usepackage{abstract} 

\usepackage{titlesec} 
\renewcommand\thesection{\Roman{section}} 
\titleformat{\section}[block]{\large\scshape\centering}{\thesection.}{1em}{} 
\titleformat{\subsection}[block]{}{\thesubsection.}{1em}{} 

\usepackage{fancyhdr} 
\usepackage{titling} 

\usepackage[dvipsnames]{xcolor}
\usepackage{hyperref}

\colorlet{mylinkcolor}{Black}
\colorlet{mycitecolor}{Black}
\colorlet{myurlcolor}{Blue}
\usepackage{graphicx}
\usepackage{cite}

\hypersetup{
 linkcolor = mylinkcolor,
 citecolor = black,
 urlcolor  = myurlcolor,
 colorlinks = true,
}

\usepackage{graphicx}

\newcommand{\ORCIDDisplay}[1]{\href{http://orcid.org/\#1}{\includegraphics[scale=.10]{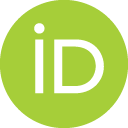}}}
\newcommand{\firstAuthor}{Gopal P. Sarma}
\newcommand{\firstAuthorEmail}{gopal.sarma@emory.edu}
\newcommand{\firstAuthorAffiliation}{School of Medicine, Emory University, Atlanta, GA USA}
\newcommand{\firstAuthorORCID}{0000-0002-9413-6202}

\pretitle{\begin{center}\huge\bfseries} 
\posttitle{\end{center}} 
\title{Scientific Literature Text Mining and the Case for Open Access} 
\author{
\textsc{
\firstAuthor\ORCIDDisplay{\firstAuthorORCID}\textsuperscript{1}\thanks{Email: \firstAuthorEmail}\hspace{2pt} 
} \vspace{8pt} \\ 
\normalsize 1. \emph{\firstAuthorAffiliation}\\ 
}
\date{} 
\renewcommand{\maketitlehookd}{%
\begin{abstract}
\noindent ``Open access'' has become a central theme of journal reform in academic publishing. In this article, I examine the relationship between open access publishing and an important infrastructural element of a modern research enterprise, \emph{scientific literature text mining}, or the use of data analytic techniques to conduct meta-analyses and investigations into the scientific corpus. I give a brief history of the open access movement, discuss novel journalistic practices, and an overview of data-driven investigation of the scientific corpus. I argue that particularly in an era where the veracity of many research studies has been called into question, scientific literature text mining should be one of the key motivations for open access publishing, not only in the basic sciences, but in the engineering and applied sciences as well. The enormous benefits of unrestricted access to the research literature should prompt scholars from all disciplines to lend their vocal support to enabling legal, wholesale access to the scientific literature as part of a data science pipeline.
\end{abstract}
}

\begin{document}

\maketitle

\section{Introduction}
The growth of institutional science following the Second World War has resulted in a range of unanticipated infrastructural problems, ranging from the overproduction of PhDs relative to the number of faculty positions, protracted educational trajectories for many aspiring scientists, and most alarmingly, a ``reproducibility crisis,'' whereby the veracity of large subsets of the research literature has been called into question \cite{Bode03061949, Bush1945, narayanamurti2013rip, Ioannidis2005, gunn2014reproducibility, adam2002journals, check2005korean, Horton2015, Alberts22042014}. \\

A significant target of institutional reform to address the larger set of issues created by dramatic scientific growth has been the academic publishing model. Access to scholarly output has traditionally been restricted to wealthy universities, whose library systems are charged exorbitant fees to maintain annual subscriptions. In contrast, an ``open access'' model is one in which research articles are made freely available to all, largely taking advantage of the infrastructural efficiencies provided by the Internet. Indeed, many new journals are online only and do not distribute printed copies of their collections. \\

Traditionally, there have been two primary arguments for open access publishing. The first is maximizing the accessibility of research output. Subscriptions to the top journals can be prohibitively expensive and only the wealthiest universities and industrial research labs can afford them. Removing the barrier to access opens the possibility for novel results to gain significantly greater exposure and scrutiny, particularly in countries with developing scientific infrastructures that have to be thrifty with resource allocation. The same is true for smaller universities or research-oriented companies for whom annual subscriptions for a full spectrum of journals cannot be reasonably budgeted. \\

The second argument for open access publishing is eliminating the ``double-billing effect'' of publicly funded research. Surely tax-paying citizens should not have to pay twice to read the output of research that they have already contributed to funding. Indeed, the Public Access of Policy of the National Institutes of Health now requires that publicly funded research be made freely available via BioMed Central within 12 months of publication \cite{NIH}. \\

An important and overlooked argument for open-access publishing is \emph{scientific literature text mining}, i.e. data science efforts which treat the scientific corpus itself as a massive dataset to analyze and extract important, actionable insights. In the remainder of the article, I give a brief discussion of scientific literature text mining, its relationship to open access publishing and the ``reproducibility crisis.'' I close with a call to all scholars to prioritize publication in journals that provide complete, unrestricted access to research articles and to support the legal availability of bulk-access to scientific papers as well as research efforts that make use of them.


\section{Scientific Literature Text Mining}
I use the phrase \emph{scientific literature text mining} to refer to data analysis of the scientific corpus, rather than the data sets that are produced by research studies. One can think of scientific literature text mining as representing a full-fledged generalization of review articles, systematic reviews, and meta-analyses whereby sophisticated tools from the modern data science toolkit are utilized to extract novel insights from the scientific corpus itself. \\

The applications of data science to the scientific corpus is in its nascent stages. It has the potential to advance our understanding of global scientific trends, the relationship between fundamental research and technological development, and fraud detection, to name just a few possible applications (see for example, \cite{markowitz2015linguistic, ding2011applying, ding2011topic, ding2013distribution, zhu2013bibliometric, zhu2015measuring, song2014productivity, valverde2007topology, gress2010properties, solee2013evolutionary})\footnote{There has been slow, but steady growth of research in recent years utilizing data science to investigate trends in the origin, growth, and dissemination of knowledge?the references given above are simply a sample of contemporary work. See, for example, research at \href{http://www.santafe.edu}{The Santa Fe Institute}, \href{http://www.knowledgelab.org}{The Knowledge Lab} at the University of Chicago, \href{http://www.nactem.ac.uk}{The National Center for Text Mining}, and the research group of \href{https://www.ch.cam.ac.uk/person/pm286}{Peter Murray-Rust} for full-fledged efforts in scientific literature text mining. See also \href{http://opencitations.net}{OpenCitations} for an open repository of citation data and \href{http://neurosynth.org}{NeuroSynth} and \href{http://neuroelectro.org}{NeuroElectro} for successfully implemented text mining platforms in the neurosciences.}. \\

Performing such analyses requires unrestricted access to the research literature so that articles can themselves can be treated as data sets to be used as part of a data science pipeline. Therefore, publishing companies which have released articles under proprietary formats, while complying with a narrow interpretation of open access, are preventing the development of a powerful set of tools and cultural practices for advancing science. Scientific literature text mining is particularly important in the context of the ``reproducibility crisis.'' Indeed, in recent years, significant attention has been drawn to low rates of reproducibility of research studies across a number of disciplines. From reproducibility initiatives, to re-examining the incentive structures of academic research, to novel journalistic practices such as post-publication peer review, the reproducibility crisis has been a significant source of controversy, discussion, and institutional action \cite{Ioannidis2005, gunn2014reproducibility, check2005korean, Horton2015, Campbell2015a, Prinz2011, economist_trouble, discover_neuroskeptic, nytimes, wired_crisis, wsj_reproducibility}. \\

However, we are barely beginning to understand the scope of the problem. For example, the studies which uncovered a large number of irreproducible results were conducted in a limited range of subjects, and we cannot generalize from these studies to understand what the ``reproducibility distribution'' looks like for the entirety of science. While discussions surrounding the reproducibility crisis have largely focused on basic research in the biological sciences, social sciences and psychology, and clinical medicine, the shared incentive structures of academic research suggests that we should view all research with a more critical eye---including the engineering and applied sciences---until the full scope of the reproducibility crisis has been more thoroughly characterized. Furthermore, as the applied sciences depend on the veracity of results producing by allied fields in basic research, these subjects are not shielded from problematic research conducted upstream in the research pipeline. Indeed, some of the most significant results in characterizing the scope of the problems in the biomedical sciences have come from the pharmaceutical industry, which relies on published drug targets as a foundational element of drug design \cite{Campbell2015a, Prinz2011}. Scientific literature text mining has the potential to play a key role in more accurately characterizing the status of different fields by identifying ``linchpin results,'' which would be of particularly high value to be the focus of targeted replication studies. For instance, by examining citation networks, it may be possible to identify a small subset of candidate papers that can be directly examined by specialists in a field to determine if adequate sample sizes or appropriate statistical tests were used. \\

The possibilities for exploratory data analysis are endless. We might imagine using natural language processing and textual analysis to characterize the transference of ideas between fields, the emergence of new concepts, and the transition from basic to applied research. Techniques such as these will also allow us to develop more refined methodologies for characterizing the importance of individual contributions and shift away from much abused metrics such as the impact factor. For example, by using predictive models or collaborative filtering, a fully digitized corpus should allow for articles to have citations automatically generated, in addition to the manually added citations by the authors themselves. These citations could be continually updated in real time as the models are refined or new research articles are published. This would allow the notion of citations to be extended to include future results as well as those that authors were aware of at the time of publication. However, none of these tantalizing possibilities can be realized if restrictions are placed on access to the research literature in bulk form.\footnote{The move towards a pre-print model for academic publications is a positive development for scientific literature text mining. Long practiced in the physics community via the arXiv pre-print server, and steadily being adopted by other fields, the pre-print model allows for drafts of publications to be immediately available online prior to submission to a journal. arXiv, originally launched by Los Alamos National Laboratory and now overseen by the Cornell University Library System, has made the full source code and PDFs of all its pre-prints available for free via Amazon Web Services \cite{arxiv_bulk}. The availability of this corpus is a powerful resource for scientific literature text mining. However, to date, it is the only pre-print server to do so and newer repositories should also follow in its footsteps. \\

As the pre-print model itself becomes more widely adopted, the simultaneous availability of both pre-prints and the final journal publication will allow for critical analysis of peer review, a facet of the modern scientific process \cite{kronick1990peer, burnham1990evolution, spier2002history} whose re-examination is essential in addressing the reproducibility crisis. For example, simple textual analysis of pre-prints and their published counterparts will allow us to characterize the extent to which peer review influences manuscripts, and how level of influence varies across different subjects.
} It is simply inadequate for articles to be released in proprietary formats that cannot be processed by automated tools. \\

Therefore, I argue that one of the primary motivations for open access publishing is the \emph{enormous benefit to science and society of scientific literature text mining}. We need a scientific analogue to CommonCrawl, an open repository of scientific articles for use in exploratory data analysis. Ironically, this argument is not new, and indeed, was anticipated as part of the original definition of open access given at the Budapest Open Access Initiative:

\begin{quote}
{\small By ``open access'' to this literature, we mean its free availability on the public internet, permitting any users to read, download, copy, distribute, print, search, or link to the full texts of these articles, crawl them for indexing, pass them as data to software, or use them for any other lawful purpose, without financial, legal, or technical barriers other than those inseparable from gaining access to the internet itself. The only constraint on reproduction and distribution, and the only role for copyright in this domain, should be to give authors control over the integrity of their work and the right to be properly acknowledged and cited \cite{budapest}.}
\end{quote}

One of the key phrases in this definition is allowing users to ``pass the articles as data to software,'' that is, \emph{to treat the scientific corpus itself as data.} However, when these words were first written, the phrase ``data science'' had yet to be coined, and the enormous growth of the field, largely driven by social media, had yet to take place. The original framers of the definition of open access had the vision and foresight to anticipate that unrestricted access to the research literature should include far more than the ability for individuals to freely read scholarly articles. They should also be able to conduct sophisticated analyses of large bodies of literature using computational techniques that have only become possible in recent years. 

\vspace{-1.5mm}

\section{Conclusion}
Realizing the full vision of scientific literature text mining requires unrestricted access to the scientific corpus. We should aspire to build a fully open infrastructure where there are APIs for every journal and pre-print repository, allowing anyone to access the data and metadata for every article and conduct exploratory or targeted data analyses. Taking advantage of a fully digitized and easily accessible corpus of knowledge, data scientists will build information dashboards providing intuitive insights into an increasingly complex knowledge base. Most importantly, scientific literature text mining may come to play a crucial role in addressing the ``reproducibility crisis'' by enabling analyses of large corpuses of scientific papers to uncover ``linchpin results'' which would subsequently be the focus of targeted replication efforts. \\

Scholars from all disciplines should be aware of the enormous benefit to society at large of scientific literature text mining and to prioritize publication in journals that allow for complete, unrestricted access to their articles. Furthermore, all researchers should lend their support to making the scientific literature legally available in bulk form and encourage data science efforts which advance this important and novel component of the modern research enterprise.

\section*{Acknowledgments}
I would like to thank Travis Rich and Daniel Weissman for insightful discussions and feedback on the manuscript. A special thanks to Devin Berg and the editorial staff of \emph{The Journal of Open Engineering} for starting an important and innovative journal and the reviewers Titus Brown and Shreejoy Tripathy for their feedback and willingness to participate in an open review process.

\section*{ORCID}
\makebox[2.5cm]{\firstAuthor} \raisebox{-.26\height}{\includegraphics[scale=.10]{orcid128}} \href{http://orcid.org/\firstAuthorORCID}{\firstAuthorORCID}\\

\bibliographystyle{ieeetr}
\bibliography{scientific_data_science}

\end{document}